\documentclass[aps,prb,twocolumn]{revtex4}   

\usepackage{amsmath}    
\usepackage{graphicx}   
\usepackage{dcolumn}

\begin{document}

\title{Almost rolling motion: An investigation of rolling grooved cylinders}
\author{Lawrence R. Mead and Frank W. Bentrem}
 \affiliation{Department of Physics and Astronomy, University of Southern Mississippi, Hattiesburg, Mississippi 39406-5046}
\date{Received 21 July 1997; accepted 12 September 1997}

\begin{abstract}
We examine the dynamics of cylinders that are grooved to form $N$ teeth for rolling motion down an inclined plane. The grooved cylinders are experimentally found to reach a terminal velocity. This result can be explained by the inclusion of inelastic processes which occur whenever a tooth hits the surface. The fraction of the angular velocity that is lost during an inelastic collision is phenomenologically found to be proportional to $2 \sin^2 \pi/N - \alpha \sin^3 \pi/N$, and the method of least squares is used to find the constant $\alpha=0.98$. The adjusted theoretical results for the time of rolling as well as for terminal velocity are found to be in good agreement with the experimental results. \\
\copyright \textit{\footnotesize{ 1998 American Association of Physics Teachers. This article may be downloaded for personal use only. Any other use requires prior permission of the author and the American Association of Physics Teachers.}}
\end{abstract}

\maketitle

\section{Introduction}

\begin{figure}
\begin{center}
\includegraphics[width=2.5in]{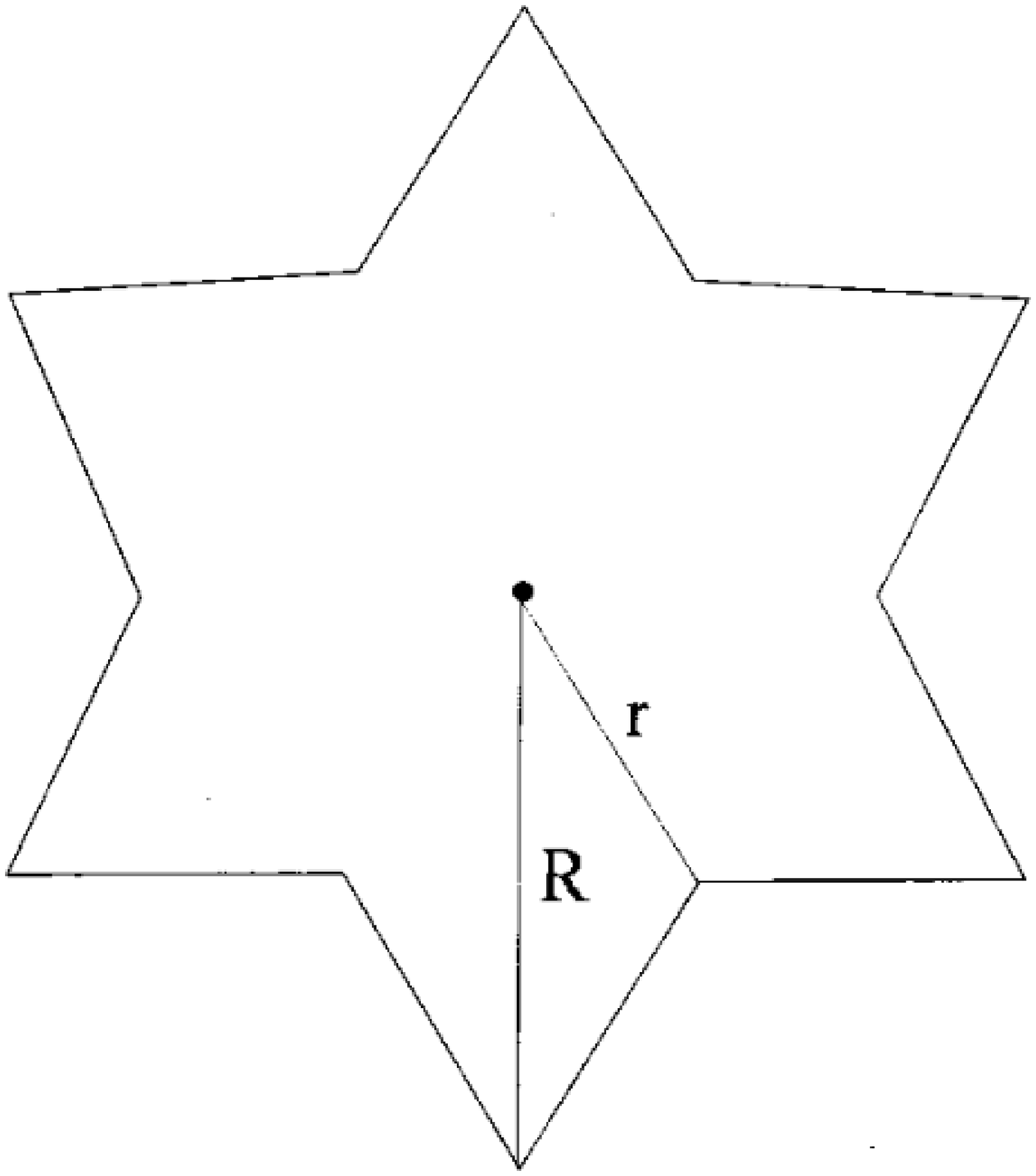}
\caption{\label{fig:fig1}A grooved cylinder with $N=6$ teeth.}
\end{center}
\end{figure}

A few years ago an article appeared in this journal \cite{lima93} which described how the fractal dimension of a randomly crumpled surface might be determined by rolling it down an inclined plane and measuring its moment of inertia $I$. In this paper, the claim is made that the moment of inertia of a hypersphere is given by 
\begin{equation}
I=\gamma MR^2,
\end{equation}
where $\gamma$ is a constant that depends on the dimension of the hypersphere, $M$ is its mass, and $R$ is its radius. This conjecture was used to attempt to measure the fractal dimension of tightly crumpled aluminum foil balls. However, it has been shown \cite{mead95} that this same sort of variation in the dynamically measured moment of inertia occurs with symmetrically grooved cylinders, such as the one depicted in Fig.~\ref{fig:fig1}, which are not fractals. 

These observations prompt one to ask to what extent the ``almost rolling'' motion of the grooved cylinders can be analyzed and understood. In particular, can the motion of regular grooved cylinders be accurately described in any simple way by dynamics and conservation laws of basic physics? Furthermore, there does not seem to be any reference to this kind of ``nearly rolling'' motion in the physics literature. Advanced engineering texts \cite{johnson85,wilson83} discuss rolling motion and contact, collisions, enmeshed gears, etc. However, the ``rolling'' of a ``gear'' or ``slotted cylinder'' does not seem to have been studied. This ``rolling'' motion of a grooved cylinder is similar to the motion of a smooth cylinder rolling on a rough surface. A rough surface tends to slow down a rolling smooth cylinder and is considered part of rolling friction,\cite{wilson83} although the dynamics of this effect has not been well analyzed. 

In this paper, we report a study of the ``nearly rolling'' motion of regularly grooved cylinders down an incline of fixed angle. In Sec.~II, the motion of such objects is analyzed using Lagrangian dynamics and making the simplifying assumption that no energy is dissipated by the ``collisions'' between the teeth and the inclined plane, but that the total energy of the cylinder is completely conserved. A recursive scheme is then derived for predicting how long a cylinder will take to roll a given, but variable, distance along an incline. Computer calculations yield predictions for these rolling times. In Sec.~III, experiments are described which were used to measure the rolling times and to compare with the theoretical predictions. The results of the first of these measurements are in marked disagreement with the theoretical predictions, with the assumption that the ``collisions'' are completely elastic. Indeed, the experimental data show that the cylinders reach terminal velocity, in contrast to the theoretical predictions.

A more precise theoretical model of this ``near rolling'' motion is then presented in Sec.~IV, which allows for inelastic processes to occur as the cylinder ``rolls.'' Predictions of this modified theory for the measured rolling times and terminal velocities are remarkably accurate.

\section{Dynamics}

Our problem consists of calculating the time it takes for a grooved cylinder to rotate about one tooth as it is rolling down an inclined plane. This section contains a discussion of the dynamics for this motion and a derivation of an expression for the time of rolling.

\subsection{Rotational kinetic energy}

We will make the assumption that the grooved cylinder is rolling without slipping, i.e., the single pivot point, $P$ in Fig.~2, does not slide. So as shown in Fig.~2, the cylinder will simply pivot about each tooth at the point $P$ until the next tooth comes in contact with the inclined plane. In Fig.~2, $\psi$ is the elevation angle of the incline and $\theta$ is the angle through which the center of mass has rotated starting from a vertical position. The kinetic energy $T$ of the rolling grooved cylinder is entirely rotational (about $P$) and is given by
\begin{equation}
T=\frac{1}{2}I_P\dot\theta^2,
\end{equation}
where $I_P$ is the moment of inertia of the grooved cylinder about the edge of the tooth in contact with the inclined plane (point $P$), and $\dot\theta$ is the angular velocity about $P$.

\begin{figure}
\begin{center}
\includegraphics[width=3.375in]{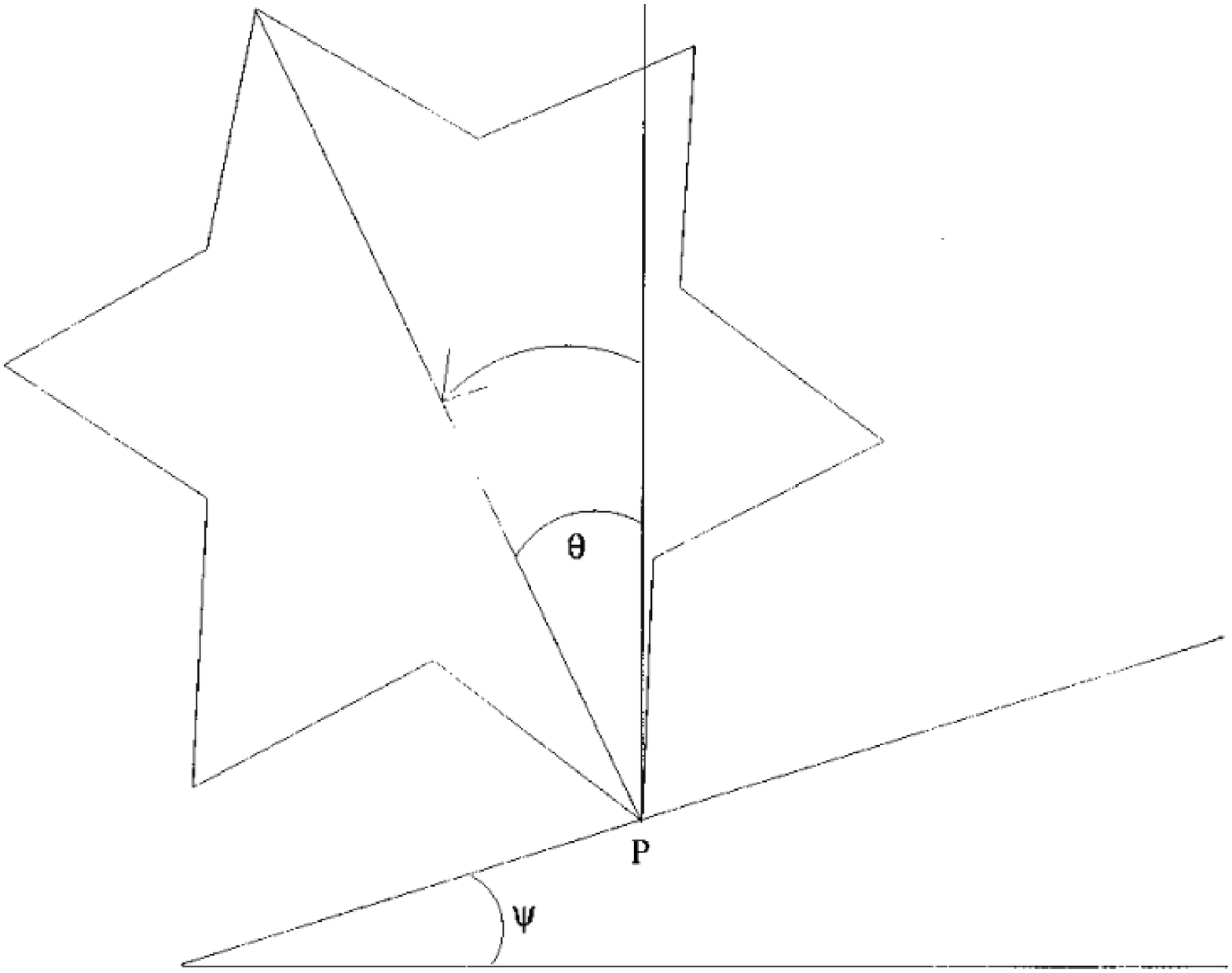}
\caption{\label{fig:fig2}A grooved cylinder with $N=6$ teeth.}
\end{center}
\end{figure}

\subsection{Time of rolling of the grooved cylinder}

 Using the parallel axis theorem, Eq.~(2) becomes
\begin{equation}
T=\frac{1}{2}\left(MR^2+I_{\text{cm}}\right)\dot\theta^2,
\end{equation}
where $M$ is the mass of the grooved cylinder, $R$ is the outer radius (i.e., the distance from the center of mass to the edge of a tooth), and $I_{\text{cm}}$ is its moment of inertia about an axis through the center of mass. If we take the potential energy to be zero when $\theta$ is zero, then the potential energy $U$ can be expressed as
\begin{equation}
U=Mgh=MgR(\cos\theta-1),
\end{equation}
using the fact that the change in height $h$ of the center of mass as a function of $\theta$ is $h=R(\cos\theta-1)$. So the Lagrangian $L$ is
\begin{equation}
L=\frac{1}{2}\left(MR^2+I_{\text{cm}}\right)\dot\theta^2+MgR(1-\cos\theta).
\end{equation}
For our problem Lagrange's equation is
\begin{equation}
\frac{d}{dt}\frac{\partial L}{\partial\dot\theta}-\frac{\partial L}{\partial\theta}=0.
\end{equation}
We can solve Lagrange's equation to obtain the result
\begin{equation}
\ddot\theta=\frac{MgR\sin\theta}{MR^2+I_{\text{cm}}},
\end{equation}
where $\ddot\theta$ is the angular acceleration about point $P$. Multiplying both sides of Eq.~(7) by $\dot\theta$ and integrating over time leads to
\begin{equation}
\dot\theta^2=C-\xi\cos\theta,
\end{equation}
where $C$ is the integration constant and $\xi$ is defined as $2MgR/(MR^2+I_{\text{cm}})$. This is a separable differential equation and can be solved to find the time $t$ it takes for the grooved cylinder to rotate about a given tooth, starting at the instant the previous tooth leaves the surface of the inclined plane and ending with the instant the next tooth strikes the surface. The result is
\begin{equation}
t=\int_{\theta_0}^{\theta_f}\frac{d\theta}{\sqrt{C-\xi\cos\theta}},
\end{equation}
where $\theta_0$ is the initial value for the angle $\theta$ and $\theta_f$ is the final angle. This is an elliptic integral and must be evaluated numerically. We find by geometry that the initial and final angles are $\psi+\pi/N$ and $\psi-\pi/N$, respectively, where $\psi$ is the elevation angle of the inclined plane (see Fig.~3). So we have
\begin{equation}
t=\int_{\psi-\pi/N}^{\psi+\pi/N}\frac{d\theta}{\sqrt{C-\xi\cos\theta}}.
\end{equation}
To solve for the integration constant, evaluate Eq.~(8) at $t=0$ and rearrange it to give
\begin{equation}
C=\omega_0^2+\xi\cos\left(\psi-\frac{\pi}{N}\right),
\end{equation}
where we have used the initial condition that $\dot\theta_0$ is the initial angular velocity $\omega_0$ when $\theta=\psi-\pi/N$. Similarly, we know that $\dot\theta$ is the final angular velocity $\omega_f$ when $\theta=\psi+\pi/N$, so that Eq.~(8) becomes
\begin{equation}
\omega_f^2=C=\xi\cos\left(\psi+\frac{\pi}{N}\right).
\end{equation}
Since for the moment we assume no loss of kinetic energy from the teeth striking the incline, the final angular velocity for the rotation about one tooth is the initial angular velocity for the rotation about the next tooth. Substitute Eq.~(12) into Eq.~(11) and rearrange to get the recursion relation
\begin{equation}
C_k=C_{k-1}+\xi\left[\cos\left(\psi-\frac{\pi}{N}\right)-\cos\left(\psi+\frac{\pi}{N}\right)\right],
\end{equation}
where $C_k$ is the integration constant for the $k$th rotation about a tooth and $C_{k-1}$ is the integration constant for the $k-1$th rotation. We can determine $C_1$ by substituting into Eq.~(11) the value of the initial angular velocity $\omega_0$ for $t=0$. 

\begin{figure}
\begin{center}
\includegraphics[width=3.375in]{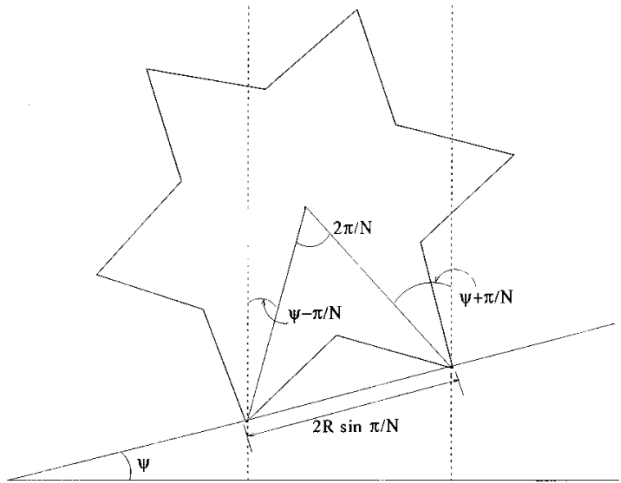}
\caption{\label{fig:fig3}The initial and final angles for rotation about a tooth.}
\end{center}
\end{figure}

The total time $\tau$ it takes for the grooved cylinder to roll a given distance is just the sum over the time integrals from Eq.~(10) for each rotation about a tooth. We write
\begin{equation}
\tau=\sum_{k=1}^nt_k,
\end{equation}
where $n$ is the total number of rotations and $t_k$ is the time integral for the $k$th rotation. It can be seen by geometry that
\begin{equation}
n=\left\lfloor\frac{L}{2R\sin(\pi/N)}\right\rfloor,
\end{equation}
where $L$ is the distance traveled along the inclined plane. We need $n$ to be an integer so the greatest integer function is used to eliminate any partial rotation at the end of the track. So we are left with the computation of
\begin{equation}
\tau=\sum_{k=1}^n\int_{\psi-\pi/N}^{\psi+\pi/N}\frac{d\theta}{\sqrt{C_k-\xi\cos\theta}}
\end{equation}
for the total time, where $C_k$ is given by Eq.~(13). 

\subsection{Moment of inertia} 

The moment of inertia about the center of mass $I_{\text{cm}}$ is calculated by breaking up the grooved cylinder into triangles as shown in Fig.~4 and performing the integration over each triangle and summing, so that 
\begin{equation}
I_{\text{cm}}=\sum_{\text{all }\Delta\text{'s}}l\rho\int dx\int dy\left(x^2+y^2\right),
\end{equation}
where $l$ is the length of the cylinder and $\rho$ is the volume density. The volume density is defined as the mass per unit volume, so we find that
\begin{equation}
\rho=\frac{M}{lNrR\sin(\pi/N)},
\end{equation}
where $r$ is the distance from the center of mass to the inside cut of the grooves (see Fig.~1). So Eq.~(17) becomes 
\begin{equation}
I_\text{cm}=\sum_{\text{all }\Delta\text{'s}}\frac{M}{NrR\sin(\pi/N)}\int dx\int dy\left(x^2+y^2\right).
\end{equation}
The result for the moment of inertia is
\begin{align}
I_{\text{cm}} = &\frac{Mr^3\cos(\pi/N)}{6R}\left(2+\cos\frac{2\pi}{N}\right) \notag \\
&+ \frac{2M}{R(R-r\cos(\pi/N))}\biggl\{\frac{r^2\sin^2(\pi/N)}{3(R-r\cos(\pi/N))^2} \notag \\
&\times \biggl[R^3\left(R=r\cos\frac{\pi}{N}\right)-\frac{3R^2}{2}\left(R^2-r^2\cos^2\frac{\pi}{N}\right) \notag \\
&+R\left(R^3-r^3\cos^3\frac{\pi}{N}\right)-\frac{1}{4}\left(R^4-r^4\cos^4\frac{\pi}{N}\right)\biggr] \notag \\
&+\frac{R}{3}\left(R^3-r^3\cos^3\frac{\pi}{N}\right)-\frac{1}{4}\left(R^4-r^4\cos^4\frac{\pi}{N}\right)\biggr\}. 
\end{align}
That this result for $I_{\text{cm}}$ yields the correct value of $\frac{1}{2}MR^2$ for a smooth cylinder can be verified by taking $N\rightarrow\infty$ and then setting $r=R$. Notice that when we substitute Eq.~(20) into Eq.~(16) the mass cancels out [substitute Eq.~(20) into $\xi$]. So the time of rolling is independent of the mass of the grooved cylinder just as it is for a perfect cylinder. 

\begin{figure}
\begin{center}
\includegraphics[width=2.5in]{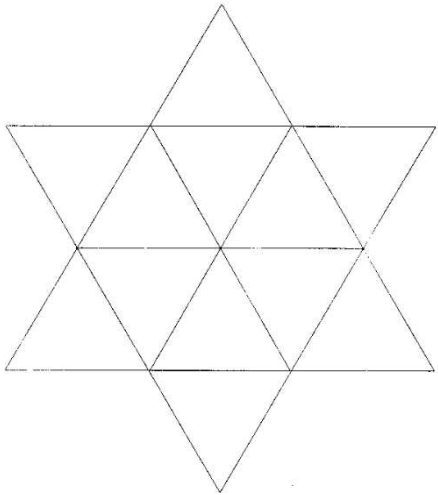}
\caption{\label{fig:fig4}A grooved cylinder divided into triangles for the purpose of calculating the moment of inertia.}
\end{center}
\end{figure}

\subsection{Computer model}

A computer program was written to numerically evaluate the integrals in Eq.~(16). The first term in the sum in Eq.~(16) contains $C_1$ , which can be calculated from Eq.~(11) for an initial angular velocity. If the grooved cylinder is started from rest with two teeth in contact with the surface, then there is some critical angle of elevation $\psi_c$ below which the cylinder will not roll. This critical angle depends on the number of teeth and is given by
\begin{equation}
\psi_c=\frac{\pi}{N}.
\end{equation}
To ensure that all the grooved cylinders will be able to roll, we let the initial angular velocity be the angular velocity that the cylinder would have if it started from rest at a balance point $\theta=0$. This corresponds to balancing a grooved cylinder and starting the timer when the first tooth strikes the surface of the inclined plane. We use conservation of energy
\begin{equation}
\Delta T=-\Delta U
\end{equation}
to find the angular velocity, where $\Delta T$ is the change in kinetic energy and $\Delta U$ is the change in potential energy. We have
\begin{equation}
\frac{1}{2}\left(MR^2+I_{\text{cm}}\right)\left(\dot\theta_f^2-\dot\theta_i^2\right)=MgR\left(\cos\theta_i-\cos\theta_f\right),
\end{equation}
where $\dot\theta_i$ and $\dot\theta_f$ are the initial and final angular velocities, and $\theta_i$ and $\theta_f$ are the initial and final angles, respectively. The final angle and final angular velocity correspond to the instant the timer is started. So we can use the final angular velocity as the initial angular velocity for the rotation about the first tooth. Now Eq.~(11) is used to find the first integration constant $C_1$. All of the other integration constants can be found by using the recursion relation given by Eq.~(13). We now have everything that we need in order to compute the time of rolling for a given grooved cylinder rolling a prescribed distance down an inclined plane. The computer model can be easily verified with a smooth cylinder simulation, i.e., take a large value for $N$ and take $r\approx R$. (We cannot have $r=R$ because this would lead to division by zero in the calculation of the moment of inertia). If we let a smooth cylinder roll a distance of $2\pi$ cm at an elevation angle of $30^\circ$, we find the analytical value for the time of rolling to be 0.196 134 s. For the computer calculation we use the values $R=1$ cm, $r=0.999$ 999 99 cm, $N=100 000$ teeth, and $L=2\pi R=6.283$ 185 cm. The numerically computed value for the time of rolling is 0.196 123 s, which is accurate to five digits.

\section{EXPERIMENTAL PROCEDURE} 

Several grooved cylinders with various numbers of teeth were needed in order to find out how the number of teeth affects the rolling motion. We used five solid aluminum cylinders which had been cut to have 12, 18, 30, 45, and 60 teeth. Table~I records the mass, inside and outside radii, and moment of inertia for these grooved cylinders. The moment of inertia was calculated using Eq.~(20). The grooved cylinders were 2.000 cm long and had an outside radius $R$ of 1.000 cm. The rolling surface used was a sheet of Plexiglas, which was rigidly attached to a thick flat board. The elevation angle for this experiment was set at $\psi=1.39\pm0.02^\circ$. An elevation angle that is too small (i.e., less than $1^\circ$) prevents the 12-tooth cylinder from proceeding down the inclined plane when released from rest. On the other hand, if an angle greater than about $3^\circ$ is used, the support points for the cylinders can actually leave the surface due to bouncing. 

\begin{table}
\begin{center}
\caption{\label{tab:tab1}Intrinsic data for the grooved cylinders.}
\begin{ruledtabular}
\begin{tabular}{ccccc}
$N$ & $M$ (g) & $R$ (cm) & $r$ (cm) & $I_{\text{cm}}\left(\text{ g } \text{cm}^2\right)$ \\
\hline
12 & 9.726 & 1.000 & 0.575  & 3.06 \\
18 & 12.860 & 1.000 & 0.735 & 4.85 \\
30 & 14.090 & 1.000 & 0.845 & 6.00 \\
45 & 14.090 & 1.000 & 0.900 & 6.72 \\
60 & 15.425 & 1.000 & 0.945 & 7.29 \\
\end{tabular}
\end{ruledtabular}
\end{center}
\end{table}

In order to measure the time of rolling a model ME-9215A Pasco Scientific Photogate Timer with Memory was attached to the board with one photogate positioned at the point where the cylinders would start to roll and another photogate whose position could be adjusted in 10-cm increments to a given distance from the first photogate. The exact distance between the photogate beams can be obtained by sliding a strip of cardboard along the incline until it ``trips'' the first photogate and marking this position. Sliding the cardboard further until it ``trips'' the second photogate and marking this position allows one to measure the distance between the marks, which is the distance between the photogate beams. 

It was observed that the rolling grooved cylinders would eventually reach terminal velocity. A microphone was plugged into a model 54600A 100-MHz, two-channel Hewlett–Packard digital oscilloscope with memory in order to pick up the clicking sound of the teeth of the rolling grooved cylinders as they hit the Plexiglas surface. Each click of a tooth created a spike on the oscilloscope. At terminal velocity the clicking sound had a constant frequency, which yielded a constant spacing between spikes on the oscilloscope. The grooved cylinders were rolled a given distance at which point the screen on the oscilloscope was frozen. Two vertical cursors were positioned on the peaks of two distinct spikes. The oscilloscope indicated the period between the two cursors. The period was divided by the distance between the teeth to get the velocity. This method for measuring the velocity of the rolling cylinders worked quite well for the 12-, 18-, and 30-toothed cylinders. However, when this technique was employed for the 45- and 60-toothed cylinders, it was difficult to separate the impact signal from the noise and echoes. In order to measure the terminal velocity for these cylinders two photogates were placed about 10-cm apart at the place the velocity was to be measured. The photogates would measure the time it took for the cylinders to roll from one gate to the other. The distance between the photogates divided by the time was the velocity. That the cylinders reached terminal velocity was verified by measuring the velocities at different rolling distances. 

In the first part of the experiment the time was measured for the grooved cylinders to roll 80.02 cm. Each cylinder was given five trials. The difficulty in this experiment was in rolling the cylinders through the narrow photogates (6.6 cm wide). These times are compared in Table~II with the times predicted from the computer program based on the Lagrangian mechanics. As one can plainly see, there is little agreement between the theoretical and experimental values. The time values are closest for the 60-toothed cylinder. As the number of teeth is decreased, the theoretical values slowly decrease but the experimental values increase significantly. In fact, the cylinders were observed to reach a terminal velocity which was lower--not higher--for fewer teeth.

\begin{table}
\begin{center}
\caption{\label{tab:tab2}Time for rolling 80 cm (energy conservation model).}
\begin{ruledtabular}
\begin{tabular}{ccc}
No. Teeth & Theoretical time (s) & Measured time (s) \\
\hline
12 & $2.74\pm0.12$ & $10.40\pm0.07$ \\
18 & $2.86\pm0.08$ & $7.29\pm0.05$ \\
30 & $2.95\pm0.08$ & $5.70\pm0.03$ \\
45 & $3.01\pm0.06$ & $5.00\pm0.03$ \\
60 & $3.05\pm0.06$ & $4.57\pm0.08$ \\
\end{tabular}
\end{ruledtabular}
\end{center}
\end{table}

The standard deviation is given as the estimated error for the measured times. Errors in the theoretical times are due to the uncertainty in the elevation angle and the final rotation about the last tooth and will be discussed later.

\section{A MORE COMPLETE THEORETICAL MODEL} 

In constructing the theoretical model of Sec.~II, two assumptions were used: that the cylinders are rotating about their point of support without slipping and that the loss of kinetic energy caused by the teeth striking the surface of the inclined plane is negligible. The first assumption is justified by the small angle of elevation; therefore, the loss of kinetic energy caused by the inelastic collisions apparently cannot be neglected.

\subsection{Angular momentum model}

In retrospect, that the collisions are inelastic should not be surprising since each tooth strikes the surface and then does not rebound but stays at that point to act as a pivot. The duration of the collision extends from the moment a tooth strikes the surface to the moment the previous tooth lifts off the surface. Conservation of angular momentum can be used to find out how much energy is lost in these collisions; for now we will assume an instantaneous collision. Choose a point on the incline where the front tooth is about to hit (just before the collision) as the origin. In general, the angular momentum $\mathbf{L}$ of the cylinder about that point is
\begin{equation} 
{\bf L}=m{\bf r}\times{\bf v}_{\text{cm}}+I_{\text{cm}}\boldsymbol{\omega}, 
\end{equation}
where $m$ is the mass of the cylinder, ${\bf r}$ is the position of the center of mass relative to that origin, and $\boldsymbol{\omega}$ is the angular velocity about the center of mass. We note with the aid of Fig.~5 that the initial velocity vector, which has the magnitude $v_i=R\dot\theta_i$, makes angle $\pi/2-2\pi/N$ with ${\bf r}_i$. Thus the initial angular momentum just before the collision is
\begin{equation}
L_i=MR\cos\left(\frac{2\pi}{N}\right)v_\text{cm}+I_\text{cm}\dot\theta_i.
\end{equation}
Similarly, the angle between ${\bf v}_f$ and ${\bf r}$ is $90^\circ$ just after the collision; hence,
\begin{equation}
L_f=\left(MR^2+I_\text{cm}\right)\dot\theta_f.
\end{equation}
In either case, $v=R\dot\theta$. By conservation of angular momentum we equate Eqs.~(25) and (26) and rearrange so that
\begin{equation}
\dot\theta_f=\left(1-\frac{2}{1+\chi}\sin^2\frac{\pi}{N}\right)\dot\theta_i,
\end{equation}
where $\chi$ is defined as $I_\text{cm}/\left(MR^2\right)$. The initial angular velocity $\dot\theta_i$ just before the collision is the final angular velocity $\dot\theta_f$ after the partial rotation given by Eq.~(12). So upon substitution we get
\begin{equation}
\dot\theta_f^2=\left(1-\frac{2}{1+\chi}\sin^2\frac{\pi}{N}\right)^2\left[C-\xi\cos\left(\psi+\frac{\pi}{N}\right)\right],
\end{equation}
where $\xi$ has been defined previously to be $2MgR/\left(MR^2+I_\text{cm}\right)$. Since the final angular velocity $\dot\theta_f$ for a rotation about one tooth is the initial angular velocity $\dot\theta_0$ for the rotation about the next tooth, Eq.~(11) becomes
\begin{align}
C_k=&\left(1-\frac{2}{1+\chi}\sin^2\frac{\pi}{N}\right)^2\left[C_{k-1}-\xi\cos\left(\psi+\frac{\pi}{N}\right)\right] \notag \\
&+\xi\cos\left(\psi-\frac{\pi}{N}\right).
\end{align}
Using this recursion relation in the computer program to find the integration constant yields the results given in Table~III. 

\begin{figure}
\begin{center}
\includegraphics[width=3.375in]{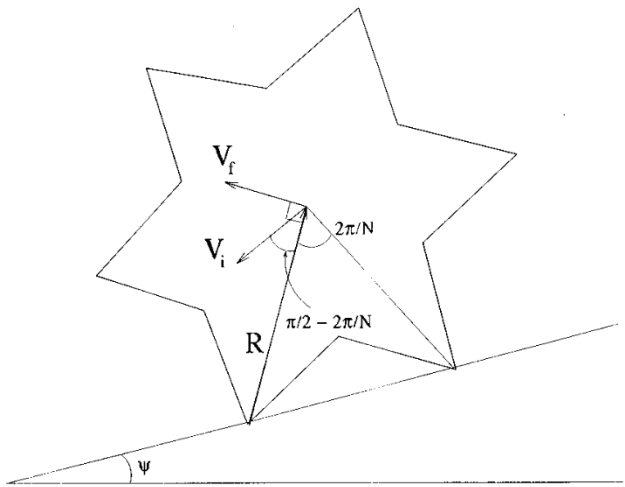}
\caption{\label{fig:fig5}Position and velocity vectors before and after the collision of a tooth with the inclined plane.}
\end{center}
\end{figure}

\begin{table}
\begin{center}
\caption{\label{tab:tab3}Time for rolling 80 cm (angular momentum model).}
\begin{ruledtabular}
\begin{tabular}{ccc}
No. Teeth & Theoretical time (s) & Measured time (s) \\
\hline
12 & $11.58\pm0.19$ & $10.40\pm0.07$ \\
18 & $7.72\pm0.10$ & $7.29\pm0.05$ \\
30 & $5.95\pm0.09$ & $5.70\pm0.03$ \\
45 & $5.09\pm0.07$ & $5.00\pm0.03$ \\
60 & $4.64\pm0.07$ & $4.57\pm0.08$ \\
\end{tabular}
\end{ruledtabular}
\end{center}
\end{table}

\subsection{Model with finite-time collisions} 

These results are somewhat better than the previous calculations. However, since we really do not have instantaneous collisions some angular momentum must be added back to the cylinder after each collision due to a net external torque which acts over a small but finite period of time. During this time the surface of the inclined plane is slightly deformed at the point of contact for both teeth, as depicted in Fig.~6. The teeth themselves can also be deformed. Figure~6 shows the relevant forces that cause the net torque about the center of mass. The forces labeled $N_1$ and $N_2$ are normal forces, and those labeled $f_1$ and $f_2$ are due to friction. Without knowing the detailed dynamics of the collision it is natural to try to find a phenomenological term that adds the correct amount to the angular momentum. 

\begin{figure}
\begin{center}
\includegraphics[width=3.375in]{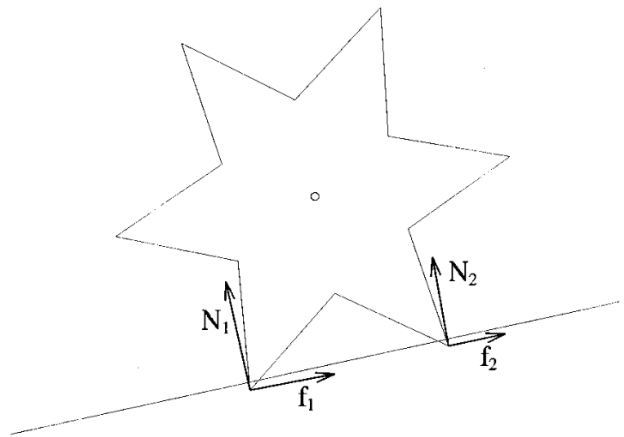}
\caption{\label{fig:fig6}Forces causing a net torque about the center of mass of the grooved cylinder.}
\end{center}
\end{figure}

If an additive term can be found which fits the measurements for 80-cm rolling times, then using this additive term we can compare the theoretical and measured results for the time of rolling at different lengths and terminal velocity. The 80-cm rolling times were chosen for the fit because of the relatively low errors in the measured times. A good fit was obtained using a term that is proportional to $\sin^3(\pi/N)$. The proportionality constant $\alpha$ was found by using the method of least squares to be $\alpha=0.98\pm0.05$. Equation~(27) becomes
\begin{equation}
\dot\theta_f=\left(1-\frac{2}{1+\chi}\sin^2\frac{\pi}{N}+\frac{\alpha}{1+\chi}\sin^3\frac{\pi}{N}\right)\dot\theta_i.
\end{equation}
This leads to the new recursion relation for the integration constant
\begin{align}
C_k=&\left(1-\frac{2}{1+\chi}\sin^2\frac{\pi}{N}+\frac{\alpha}{1+\chi}\sin^3\frac{\pi}{N}\right)^2 \notag \\
&\times\left[C_{k-1}-\xi\cos\left(\psi+\frac{\pi}{N}\right)\right]+\xi\cos\left(\psi-\frac{\pi}{N}\right).
\end{align}
Table IV compares the theoretical and measured rolling times for 80 cm using this additional angular momentum term. 

\begin{table}
\begin{center}
\caption{\label{tab:tab4}Time for rolling 80 cm (model with finite-time collisions).}
\begin{ruledtabular}
\begin{tabular}{ccc}
No. Teeth & Theoretical time (s) & Measured time (s) \\
\hline
12 & $10.35\pm0.19$ & $10.40\pm0.07$ \\
18 & $7.40\pm0.12$ & $7.29\pm0.05$ \\
30 & $5.84\pm0.11$ & $5.70\pm0.03$ \\
45 & $5.04\pm0.09$ & $5.00\pm0.03$ \\
60 & $4.61\pm0.08$ & $4.57\pm0.08$ \\
\end{tabular}
\end{ruledtabular}
\end{center}
\end{table}

An accurate description of the ``rolling'' motion of grooved cylinders, in addition to predicting rolling times, should be able to predict the terminal velocities of the cylinders. Experiment showed that each of our grooved cylinders had reached terminal velocity after rolling 60 cm. The velocities of the cylinders were measured after rolling 115 cm and, taking an average over ten trials, were compared to the predictions from our modified theory. Table~V lists these values. 

\begin{table}
\begin{center}
\caption{\label{tab:tab5}Terminal velocity.}
\begin{ruledtabular}
\begin{tabular}{ccc}
No. Teeth & Theoretical velocity & Measured velocity \\
& (cm/s) & (cm/s) \\
\hline
12 & $7.92\pm0.10$ & $8.08\pm0.26$ \\
18 & $11.43\pm0.10$ & $11.78\pm0.42$ \\
30 & $15.23\pm0.12$ & $15.25\pm0.10$ \\
45 & $18.67\pm0.14$ & $18.46\pm0.15$ \\
60 & $21.52\pm0.16$ & $21.57\pm0.24$ \\
\end{tabular}
\end{ruledtabular}
\end{center}
\end{table}

Rolling times were measured for several different lengths. The results are given in Table~VI. Table~VII contains the predicted rolling times for the lengths used in Table VI. These predictions come from the modified theory which includes inelastic processes. The proportionality constant $\alpha=0.98$ comes from the least-squares best fit to the 80-cm rolling times. The agreement between the measured and predicted results is remarkable considering we used only a single best-fit parameter. The only measured value which did not fall within the error estimates of the predicted value is the 40-cm rolling time for the 12-toothed cylinder; the most difficult case from which to obtain reliable results.

\begin{table}
\begin{center}
\caption{\label{tab:tab6}Measured rolling times at various lengths.}
\begin{ruledtabular}
\begin{tabular}{ccccc}
No. Teeth & 20 cm (s) & 40 cm (s) & 60 cm (s) & 80 cm (s) \\
\hline
12 & $2.98\pm0.09$ & $5.60\pm0.05$ & \dots\cite{dataNote} & $10.40\pm0.07$ \\
18 & $2.19\pm0.05$ & $3.93\pm0.05$ & \dots\cite{dataNote} & $7.29\pm0.05$ \\
30 & $1.95\pm0.05$ & $3.19\pm0.02$ & $4.53\pm0.06$ & $5.70\pm0.03$ \\
45 & $1.82\pm0.08$ & $2.89\pm0.05$ & $4.02\pm0.06$ & $5.00\pm0.03$ \\
60 & $1.76\pm0.04$ & $2.75\pm0.04$ & $3.75\pm0.11$ & $4.57\pm0.08$ \\
\end{tabular}
\end{ruledtabular}
\end{center}
\end{table}

\begin{table}
\begin{center}
\caption{\label{tab:tab7}Theoretical rolling times at various lengths.}
\begin{ruledtabular}
\begin{tabular}{ccccc}
No. Teeth & 20 cm (s) & 40 cm (s) & 60 cm (s) & 80 cm (s) \\
\hline
12 & $2.77\pm0.12$ & $5.32\pm0.13$ & $7.80\pm0.15$ & $10.35\pm0.17$ \\
18 & $2.14\pm0.10$ & $3.91\pm0.10$ & $5.64\pm0.11$ & $7.40\pm0.12$ \\
30 & $1.88\pm0.10$ & $3.21\pm0.10$ & $4.53\pm0.10$ & $5.84\pm0.09$ \\
45 & $1.78\pm0.08$ & $2.89\pm0.09$ & $3.97\pm0.09$ & $5.04\pm0.07$ \\
60 & $1.73\pm0.07$ & $2.73\pm0.07$ & $3.68\pm0.07$ & $4.61\pm0.07$ \\
\end{tabular}
\end{ruledtabular}
\end{center}
\end{table}

\section{Conclusions}

We found that the ``rolling'' motion of a symmetrically grooved cylinder can be accurately described by assuming that the cylinder rotates about each tooth without slipping. Angular momentum is nearly conserved during the collision, however, inelastic processes cause a net torque to act on the cylinder over a small but finite period of collision time. (Recall that the collision period extends from the moment the tooth comes in contact with the inclined plane to the moment the previous tooth lifts off the plane.) 

The time it takes a grooved cylinder to ``roll'' a distance $L$ along an incline with an elevation angle $\psi$ is given by Eqs.~(15) and (16). The integration constant in Eq.~(16) is given by Eq.~(31), where the proportionality constant $\alpha$ was found to be $0.98\pm0.05$ using a least-squares best fit to the 80-cm rolling times. 

This constant probably depends on the coefficient of restitution of the cylinders on the incline. The constant $\alpha$ may also depend on the elevation angle, mass and radii of the grooved cylinders, and elasticity (leading to bending moments) of the cylinders and the inclined plane. 

The estimated errors given for the measured rolling times and terminal velocities are the standard deviations of the measurements. The estimated theoretical errors come from two sources. The first error source is the uncertainty in the elevation angle. The angle was measured to an accuracy within $0.02^\circ$. The amount of error from this source was estimated by using this variation in the computer program ($1.37^\circ$ instead of $1.39^\circ$) and finding the difference from the originally computed value. Another significant error source comes from not having a complete final rotation about the last tooth. The cylinder ``trips'' the final timing gate before the final rotation is complete. The predicted times include half of the time it takes to make the final rotation which is also the estimated error from this source. 

Additional error comes from the cylinder's tendency to travel in a slightly curved path. This tendency would clearly result in measured times which are longer than for the assumed straight line path. One final source of error comes from the uncertainty in the initial angle and the initial angular velocity of the grooved cylinders. The cylinders were released from a vertical balanced position and ``tripped'' the initial timing gate just after release. It was assumed that the timer started after the cylinder had rotated $0.5^\circ$. The amount of error from the latter two sources is not likely to be large and would be difficult to estimate. 

Future study on ``rolling'' motion of grooved cylinders could focus on determining how the proportionality constant $\alpha$ depends on the various parameters mentioned above. Grooved cylinders that are not symmetrical would also be interesting to study. Finally, spheres that are ``grooved'' in three dimensions such as tightly crumpled aluminum foil balls could be investigated. This could contribute to a better understanding of rolling friction caused by surface roughness.

\begin{acknowledgments}
Special thanks go to Dr. Raymond Folse and Glen Messer for their valuable insight and suggestions throughout the experimental phase. Cooperation from the School of Engineering Technology of the University of Southern Mississippi in providing a digital oscilloscope is much appreciated. Finally, we would like to express gratitude for the funds received from the College of Science and Technology (USM) for manufacturing the grooved cylinders.
\end{acknowledgments}


\begin{thebibliography}{5}
\bibitem{dataNote}No data are available for 60-cm rolling times for the cylinders with 12 and 18 teeth because of the problem of rolling these cylinders through the narrow timing gates. The 80-cm rolling times were obtained with great difficulty.
\end{thebibliography}
\end{document}